\documentclass{aastex631}

\begin{document}

\title{Multi-Frequency General Relativistic Radiation-Magnetohydrodynamic Simulations of Thin Disks}

\author[0000-0002-5786-186X]{P. Chris Fragile}
\affiliation{Department of Physics and Astronomy, College of Charleston, 66 George Street, Charleston, SC 29424, USA}
\affil{Kavli Institute for Theoretical Physics, University of California Santa Barbara, Santa Barbara, CA 93106, USA}

\author{Peter Anninos}
\affiliation{Lawrence Livermore National Laboratory, P.O. Box 808, Livermore, CA 94550, USA}

\author[0000-0002-6485-2259]{Nathaniel Roth}
\affiliation{Lawrence Livermore National Laboratory, P.O. Box 808, Livermore, CA 94550, USA}

\author{Bhupendra Mishra}
\affiliation{Harish-Chandra Research Institute, HBNI
Chhatnag Road, Jhusi, Allahabad 211019, India}

\begin{abstract}
We present a set of six general relativistic, multi-frequency, radiation magnetohydrodynamic simulations of thin accretion disks with different target mass accretion rates around black holes with spins ranging from non-rotating to rapidly spinning. The simulations use the $\mathbf{M}_1$ closure scheme with twelve, independent frequency (or energy) bins ranging logarithmically from $5\times 10^{-3}$ to $5\times 10^3$ keV. The multi-frequency capability allows us to generate crude spectra and energy-dependent light curves directly from the simulations without a need for special post-processing. While we generally find roughly thermal spectra with peaks around 1 to 4 keV, our high-spin cases showed harder than expected tails for the soft or thermally dominant state. This leads to radiative efficiencies that are up to five times higher than expected for a Novikov-Thorne disk at the same spin. We attribute these high efficiencies to the high-energy, coronal emission. These coronae mostly occupy the effectively optically thin regions near the inner edges of the disks and also cover or sandwich the inner $\sim 15 GM/c^2$ of the disks.
\end{abstract}


\textbf{}
\section{Introduction} \label{sec:intro}
During outbursts, black hole X-ray binaries (BHXRBs) spend most of their time in one of two prominent states: a hard state, characterized by a power-law spectrum with photon index $\Gamma \lesssim 1.8$ and aperiodic variability \citep{vanderKlis06,Remillard06,Done07}; and a soft state, in which a thermal component dominates and variability is negligible \citep[e.g.][]{vanderKlis06,Done07}. A typical outburst starts in the hard state, switches to the soft state at luminosities $\gtrsim 0.6 L_\mathrm{Edd}$, where $L_\mathrm{Edd} = 1.2 \times 10^{38} (M/M_\odot)$ erg s$^{-1}$ is the Eddington luminosity of a black hole of mass $M$, and ultimately switches back to the hard state at a slightly lower luminosity $L \approx 0.2 L_\mathrm{Edd}$ \citep[showing hysteresis;][]{Miyamoto95,Maccarone03}. 

The thermal contribution to the soft-state spectrum is presumed to come from a radiatively efficient, thin disk \citep{Shakura73,Novikov73}, which will be the main focus of this paper. We hope to similarly explore other spectral states in future work. To date, these thin disks have been difficult to simulate, especially in the luminosity range relevant to BHXRBs. This is because they suffer from thermal \citep{Shakura76}, and possibly viscous \citep{Lightman74}, instabilities precisely in the luminosity range identified with the soft state, while observations show little support for such instabilities \citep[although see][and references therein for arguments that such instabilities are at play in GRS 1915+105]{Janiuk02}. Technically, these instabilities have only been formally proven for $\alpha$-viscosity disks \citep[e.g.][]{Fragile18}, although simulations have shown that turbulent, magnetohydrodynamic (MHD) disks can also be susceptible to the thermal instability \citep{Jiang13,Mishra16,Ross17}. However, some magnetic field configurations, especially those that can maintain significant magnetic pressure near the disk midplane, appear capable of stabilizing such disks \citep{Begelman07,Oda09,Sadowski16a}. One such configuration is a radially extended quadrupole field \citep{Mishra22}, which is what we use for the simulations presented in this paper.

Even in the soft, thermally dominant state, there is usually a residual hard, power-law component. Generally, this hard component is attributed to a ``corona'' of unknown origin and geometry. Three of the most commonly suggested geometries include: one in which the corona is in the shape of a torus or doughnut occupying a region between the inner edge of the thin disk and the central object, possibly overlapping somewhat with the disk \citep{Gierlinski97,Esin01}; another in which it is thought to occupy a relatively compact region directly above and below the black hole, possibly at the base of the jet \citep{Fender99,Markoff05}; and lastly in a configuration where it simply sandwiches the disk \citep{Haardt91,Haardt93}. Currently, there are compelling observational arguments for all three geometries, though they are obviously in tension with one another, as they make different predictions about how different spectral features should correlate. 

Models used to fit the soft component vary in their levels of sophistication. The simplest model \citep[DISKBB;][]{Mitsuda84} has only two free parameters, one for the inner radius of the accretion disk, $r_\mathrm{in}$, and another for the temperature at that radius, $T_\mathrm{in}$. The temperature in the rest of the disk is assumed to follow a power law, $T(r) = T_\mathrm{in} (r/r_\mathrm{in})^{-3/4}$, and the total luminosity is simply $L_\mathrm{disk} = 4\pi r_\mathrm{in}^2 \sigma T_\mathrm{in}^4$. More advanced models include additional physics considerations, such as adopting a stress–free inner boundary, which requires the temperature to drop to zero at $r_\mathrm{in}$ \citep{Gierlinski99,Kubota01}; including a spectral hardening factor to account for the incomplete thermalization of the escaping radiation \citep{Merloni00,Davis05}; and accounting for various relativistic corrections \citep{Zhang97}. For more details about modeling disk spectra, see \citet{Done07}.

Multi-frequency, radiation MHD simulations, like the ones presented in this paper, present a new way to generate disk models and corresponding spectra simultaneously. They also provide some hope for disentangling some of the issues mentioned, such as identifying the source of any hard X-rays and defining the geometry of the corona by tracing where light at different energies originates. Another benefit is that we have the ability to create independent light curves for each energy/frequency bin, which will allow us to look for correlated variability, much the way observers do \citep{Miyamoto89,Nowak99,McHardy04}. In the future, we hope to use this same method to explore the hard- and soft-intermediate states, where hard and soft photons provide nearly equal contributions to the spectrum. In those cases, multi-frequency radiation MHD capabilities may be indispensable to properly modeling all the components. 

Previous studies have created spectra and energy-dependent light curves from simulations through post-processing via a separate radiation transport code. Some of those have started from non-radiative, GRMHD simulations with an artificial cooling function \citep{Kinch19,Kinch21}; others have evolved the radiation self-consistently within the simulation, though with a grey opacity \citep{Narayan16}. Only very recently have groups started using multi-frequency radiation MHD simulations \citep{Mills23}. In a companion paper (Roth et al. in preparation), we will compare the results of the present work with our own radiation transport post-processing code \citep{Roth22}.

The rest of this paper focuses on describing the setup of our simulations (Section \ref{sec:methods}); comparing our new multi-frequency simulations with an earlier, comparable, frequency-integrated (grey) simulation (Section \ref{sec:grey}); exploring the effects of varying the target mass accretion rate (Section \ref{sec:mdot}); characterizing the resulting spectra and light curves that we get from the simulations (Section \ref{sec:spectra}); and studying the effects of black hole spin (Section \ref{sec:spin}). Finally, we wrap everything up in Section \ref{sec:discussion}. All of our simulations assume a black hole of mass $M = 6.62 M_\odot$; therefore, our distance unit, $r_G = GM/c^2$, is equal to 9.8 km, and our time unit, $t_G = GM/c^3$ is equal to $3.3 \times 10^{-5}$ s. We also assume solar composition within the disk, with a mean molecular weight of $\mu = 0.615$.

\section{Methods and Models}
\label{sec:methods}

In this paper, we present six, new, three-dimensional general relativistic, radiation magnetohydrodynamic (MHD) simulations that are intended to cover a range of mass accretion rates and black hole spins. Each simulation starts from a standard Novikov-Thorne disk configuration \citep{Novikov73,Abramowicz13}, assuming the viscosity parameter to be $\alpha = 0.02$ and an adiabatic equation of state with $\gamma=4/3$, though we emphasize that the current simulations do not employ any form of explicit viscosity. This disk is seeded with an initially weak, radially extended, quadrupole magnetic field. This field was chosen because in \citet{Mishra22} we demonstrated that it can be stable at the targeted mass accretion rates. 

\subsection{Numerical Details}

The general set up of the disk and magnetic field are described in detail in \citet{Fragile18} and \citet{Mishra22}, so we forgo that here and focus instead on the relevant numerical details. All simulations are carried out using the general relativistic, radiation, magnetohydrodynamics (GRRMHD) code, Cosmos++ \citep{Anninos05}. We use the high resolution shock capturing (HRSC) scheme described in \citet{Fragile12} to solve for the flux and gravitational source terms of the gas and radiation. Rather than evolving the magnetic fields directly, we instead evolve the vector potential and recover the fields from it as needed, as described in \citet{Fragile19}. 

For the radiation, we use the $\mathbf{M}_1$ closure scheme, which retains the first two moments of the radiation intensity and (average) radiative flux. The new aspect is that we use the recently added multi-frequency capabilities of Cosmos++ \citep{Anninos20} to evolve the radiation in twelve frequency bins rather than a single one (grey opacity) as in \citet{Mishra22}. Since part of the purpose of this paper is to compare and contrast results between the grey-opacity and multi-frequency simulations, we first highlight some of the differences between the two methods \citep[see][for more details]{Anninos20}. The first major difference is that the multi-frequency radiation conservation equation
\begin{equation}
R^{\alpha\beta}_{(\nu) ; \beta} 
   - \frac{\partial}{\partial\nu}\left[\nu M^{\alpha\beta\gamma}_{(\nu)} u_{\beta ;\gamma}\right]
   = -G_{\alpha (\nu)}   ~,
\end{equation}
includes an extra term (the second one), which captures the Doppler and gravitational frequency shifts of the radiation. Other differences are found in the radiation four-force density (coupling) term. For the grey opacity simulations, this has the form
\begin{eqnarray}
    G^{\mu} & = &-\rho (\kappa^\mathrm{a}_F+\kappa^s)R^{\mu \nu}u_{\nu}- \rho \left\{ \left[\kappa^s+4\kappa^s\left(\frac{T_{\mathrm{gas}}-T_{\mathrm{rad}}}{m_e}\right) +\kappa^\mathrm{a}_F - \kappa^\mathrm{a}_J \right]   R^{\alpha \beta}u_\alpha u_\beta +\kappa^\mathrm{a}_\mathrm{P}a_RT^4_\mathrm{gas}\right\} u^\mu ~,
\label{eqn:Gmu}
\end{eqnarray}
where $\kappa^\mathrm{a}_\mathrm{P}=2.8\times10^{23}\,T^{-7/2}_\mathrm{K} \rho_{\mathrm{cgs}}$ cm$^2$ g$^{-1}$ and $\kappa^\mathrm{a}_\mathrm{R}=7.6\times10^{21}\,T^{-7/2}_\mathrm{K}\,\rho_{\mathrm{cgs}}$ cm$^2$ g$^{-1}$ are the Planck and Rosseland mean opacities for free-free absorption, respectively, $\kappa^s = 0.34$ cm$^2$ g$^{-1}$ is the opacity due to electron scattering, $R^{\mu\nu}$ is the radiation stress tensor, $u^\mu$ is the fluid four-velocity, $T_{\mathrm{K}}$ is the ideal gas temperature of the fluid in Kelvin, and $\rho_{\mathrm{cgs}}$ is density in g cm$^{-3}$. Thus, we are assuming Kramers-type opacity laws, with the Rosseland mean also used for the flux mean, $\kappa^\mathrm{a}_\mathrm{F} = \kappa^\mathrm{a}_\mathrm{R}$, and the Planck mean used for the J-mean, $\kappa^\mathrm{a}_\mathrm{J} = \kappa^\mathrm{a}_\mathrm{P}$. This form also includes an approximate Compton cooling term (the second term in square braces). The form of the radiation four-force density used for the multi-frequency simulations,
\begin{equation}
G^\mu_{(\nu)} = -\rho \left[\kappa^{\mathrm{a}}_{(\nu)} + \kappa^{\mathrm{s}}_{(\nu)}\right] R^{\mu \nu }_{(\nu)} u_{\nu} 
                -\rho\left[\kappa^{\mathrm{s}}_{(\nu)} R^{\alpha \beta}_{(\nu)} u_{\alpha} u_{\beta} + 
                           \kappa^{\mathrm{a}}_{(\nu)} B_{(\nu)} \right] u^{\mu}~,
\label{eq:Galpha}
\end{equation}
varies in some significant ways. First, the opacities, $\kappa^\mathrm{a}_{(\nu)}$ and $\kappa^\mathrm{s}_{(\nu)}$, are now 
frequency-dependent. Second, the frequency-integrated radiation energy density, $a_RT^4_\mathrm{gas}$, is replaced by the Bose-Einstein statistical distribution function for the energy density as a function of frequency,
\begin{equation}
B_{(\nu)} = \frac{8\pi (h\nu)^3}{(hc)^3}\left(\frac{1}{e^{(h\nu)/(kT)} - 1}\right)  ~.
\end{equation}
Finally, the multi-frequency simulations use a Fokker-Planck approximation (i.e., the Kompaneets equation) to track the Compton scattering of the photons. We mention some of the potential impacts of these differences in Section \ref{sec:grey}.

For the multi-frequency simulations, the thirteen energy bin edges are spaced logarithmically between $5\times 10^{-3}$ and $5\times 10^3$ keV. This gives reasonable coverage near the emission peak ($\sim 1-5$ keV) for a stellar-mass black hole accretion disk, while keeping the overall cost of each simulation manageable. The number of bins, $N$, is constrained strongly on the high side by computational cost (finite resources). This is due primarily to the performance of the $\mathbf{M}_1$ method, which transitions from roughly linear scaling at low bin resolution to a much more costly scaling of about $N^{2.5}$ at high bin count. This transition, which occurs at $N \sim 10$ \citep{Roth22}, is attributed to the greater relative cost associated with solving the $5+4N$ dimensional nonlinear primitive inversion problem when the matrix operations begin to exceed the cost of solving the additional $4N$ transport equations. Considering the high computational demands required simply for simulating thin disks, we opted to maximize the number of bins while staying within the linear scaling range. Although not all of our simulations require as wide an energy bin range as quoted above, we opted for consistency purposes to maintain the same range for all simulations. A 5 eV low energy edge provides an accounting of energy $\sim3$ orders of magnitude below the thermal peak. An upper edge of a few MeV is required by the highest black hole spins to provide a comparable energy accounting for inverse Compton scattered photons, and to begin to observe the high energy transition from Compton to bremsstrahlung signatures attributed to high temperature regions near the black hole horizon. While the non-spinning black hole simulations do not require upper energy cutoffs this high, we kept the range the same for all simulations for consistency.

All of the simulations are carried out on a three-level, nested (statically refined) spherical polar grid with resolution concentrated near the black hole and toward the midplane of the disk. We use a logarithmic radial coordinate to cover the range from generally somewhat inside the ISCO radius out to $160 r_G$, and only consider a quarter of the azimuthal domain ($0 \le \phi \le \pi/2$). Further details of the grid setup are provided in \citet{Mishra22}. The base resolution is $48\times48\times12$, giving a peak resolution equivalent to $192\times192\times48$, with proper cell dimensions of $(\Delta r_\mathrm{p}, \Delta \theta_\mathrm{p}, \Delta \phi_\mathrm{p}) = (0.234r_G, 0.049r_G, 0.328r_G)$ at $r=10r_G$, $\theta = \pi/2$. This physical resolution is as good or better than most studies of geometrically thick accretion flows \citep[e.g.,][]{Porth19}, but for these exceptionally thin disks, it may still not be enough. For instance, in the vertical direction, we only manage seven zones per disk scale height. However, given the extra cost of multi-frequency radiation, we could not realistically do better at this point. Fortunately, when we compared (grey opacity) simulations at this same resolution with ones at twice this resolution in \citet{Mishra22}, we found the results to be largely converged. Additionally, the magneto-rotational instability (MRI) is adequately resolved according to the standard quality factors \citep{Hawley11}, as $Q^{(z)}\ge 8$ and $Q^{(\phi)} \ge 90$ when time averaged over the final 1,000$t_G$.\footnote{The azimuthal quality factor is so high because we purposefully chose a field configuration that produces a strong toroidal field component in order to try to stabilize the disk.}

\subsection{Model Descriptions}

We now describe the parameter choices for each simulation, in turn. The full collection of simulations, along with their relevant parameters, particularly the Novikov-Thorne radiative efficiency, $\eta_\mathrm{NT}$; the inner radius of simulation domain, $r_\mathrm{min}$; and the stop time for each simulation, $t_\mathrm{stop}$, are listed in Table \ref{tab:params}. The stop times range from $t_\mathrm{stop} = 6,369$ to $15,045 t_\mathrm{G}$ for the multi-frequency simulations, long enough in the latter case to achieve a steady-state solution out to roughly $20 r_G$.
\begin{deluxetable}{ccccccl}
\tablecaption{Simulation Parameters \label{tab:params}}
\tablecolumns{7}
\tablehead{
\colhead{Sim} & \colhead{$\dot{m}$} & \colhead{$a_*$} & \colhead{$\eta_\mathrm{NT}$} & \colhead{$r_\mathrm{min}/r_G$} & \colhead{$t_\mathrm{stop}/t_G$} & \colhead{Notes}}
\startdata
S3Ea0 & 3 & 0 & 0.057 & 4.0 & 15,045 & Collapsed \\
S3Ea75 & 3 & 0.75 & 0.112 & 2.6 & 6,369 & \\
S3Ea9 & 3 & 0.9 & 0.156 & 1.9 & 10,337 & \\
S3Ea95 & 3 & 0.95 & 0.190 & 4.0 & 10,430 & $r_\mathrm{min} > r_\mathrm{ISCO}$ \\
S7Ea5 & 7 & 0.5 & 0.082 & 4.0 & 8,399 & Collapsed \\
S10Ea95 & 10 & 0.95 & 0.190 &  4.0 & 10,522 & $r_\mathrm{min} > r_\mathrm{ISCO}$ \\
S3Ea0\_grey & 3 & 0 & 0.057 & 4.0 & 30,000 &
\enddata
\end{deluxetable}

The first simulation, S3Ea0, uses a target mass accretion rate of $\dot{m} = \dot{M} c^2/L_\mathrm{Edd} = 3$ onto a non-rotating, Schwarzschild black hole ($a_* = 0$). This simulation shares the same parameters and setup as the S3Eq model in \citet{Mishra22}, though we refer to that model in this paper as S3Ea0\_grey. The two S3Ea0 simulations differ only in whether they use multiple frequency bins to resolve the radiation spectrum (S3Ea0) or a single, integrated (grey) opacity (S3Ea0\_grey). With these two simulations, we attempt to assess the impact of our multi-frequency capability.

The second simulation, S3Ea75, is the first to introduce spin to the black hole ($a_* = 0.75$ in this case), which provides the possibility of forming a Blandford-Znajek jet \citep{Blandford77}. This is important, because, as mentioned, one of the suggestions in the literature is that the hard X-rays that make up the ``corona'' are, in fact, coming from the base of the jet \citep{Fender99,Markoff05}. Our simulations will be able to directly test this idea, at least if the corona can be attributed to thermal processes. It will also give us some idea of how well any X-rays from the jet are able to illuminate the disk, a key assumption in many applications of reflection modeling \citep[e.g.,][and references therein]{Reynolds14} and reverberation mapping \citep[e.g.,][and references therein]{Uttley14}. This is also where looking at cross-correlations between light curves in different energy bins and mapping where different spectral components originate will be of great use.

The third, S3Ea9, and fourth, S3Ea95, simulations increase the spin further to $a_* = 0.9$ and 0.95, respectively. The question will be whether and how this affects the hard X-ray luminosity. The higher black hole spin should lead to a more powerful jet, so if the jet is the source of the hard X-rays then there should be a clear distinction between the different simulations and we could quantify how the hard X-ray luminosity correlates with spin. However, previous simulations of black hole accretion disks and jets have shown that most of the jet power initially comes out in the form of Poynting flux and must be converted into radiation further downstream \citep{Tchekhovskoy08,Bromberg16}, so it is not clear how much that component should contribute to our spectra. Furthermore, the soft state has not been found to exhibit significant jet activity \citep{Fender99,Russell11,Maccarone20}, so it would, in fact, be somewhat surprising if we saw strong jets in our simulations. Another issue with the S3Ea95 simulation is that we accidentally left the inner domain boundary at $r_\mathrm{min} = 4 r_G$, which is outside the innermost stable circular orbit (ISCO) of the accretion disk for a black hole of this spin.

Finally, we consider a couple simulations, S7Ea5 and S10Ea95, with higher target mass accretion rates, specifically $\dot{m} = 7$ and 10, with spins of $a_*=0.5$ and 0.95, respectively. Unfortunately, model S7Ea5 ($\dot{m}=7$ and $a_*=0.5$) collapsed vertically due to the previously mentioned thermal instability and settled to a lower $\dot{m}$ branch of its thermal equilibrium curve; the scale height became under-resolved; and the disk did not recover over the duration of that simulation. This is similar to what happened to simulation S10E in \citet{Fragile18}. Therefore, we do not discuss the S7Ea5 simulation further in this paper. The S10Ea95 simulation similarly collapses initially, but unlike S7Ea5, it gradually recovers as the simulation progresses. This behavior likely at least partially reflects the fact that it takes time for the source of the turbulent heating in the disk, namely the MRI, to build up, though radiative cooling is active right from the start. In the future, it may be better to run these simulations for some time with radiative cooling inhibited while the MRI builds up. The mistake of leaving the inner domain boundary at $r_\mathrm{min} = 4 r_G$ unfortunately also applies to simulation S10Ea95.

\section{Grey vs. Multi-frequency}
\label{sec:grey}

One of our main goals in this paper was to study the differences between grey (single-frequency) and multi-frequency simulations of thin disks. This was supposed to be accomplished by comparing the S3Ea0\_grey simulation from \citet{Mishra22} with the new S3Ea0 one. Unfortunately, despite these two simulations only differing in their radiative treatments, they, nevertheless, followed different evolutionary paths. 

\begin{figure}
\centering
\includegraphics[width=0.31\textwidth]{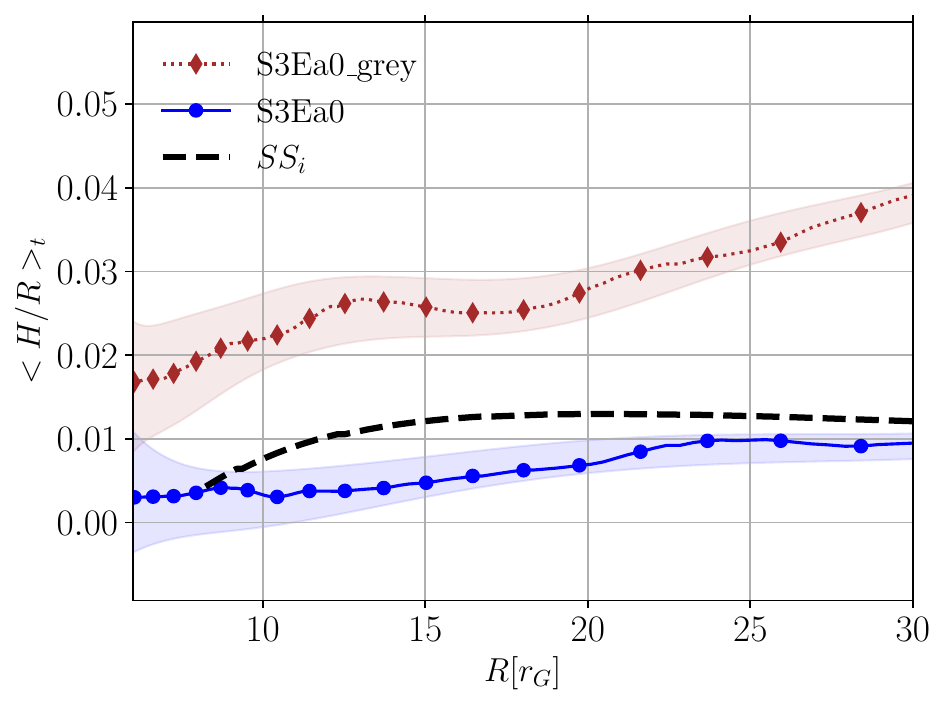}
\includegraphics[width=0.31\textwidth]{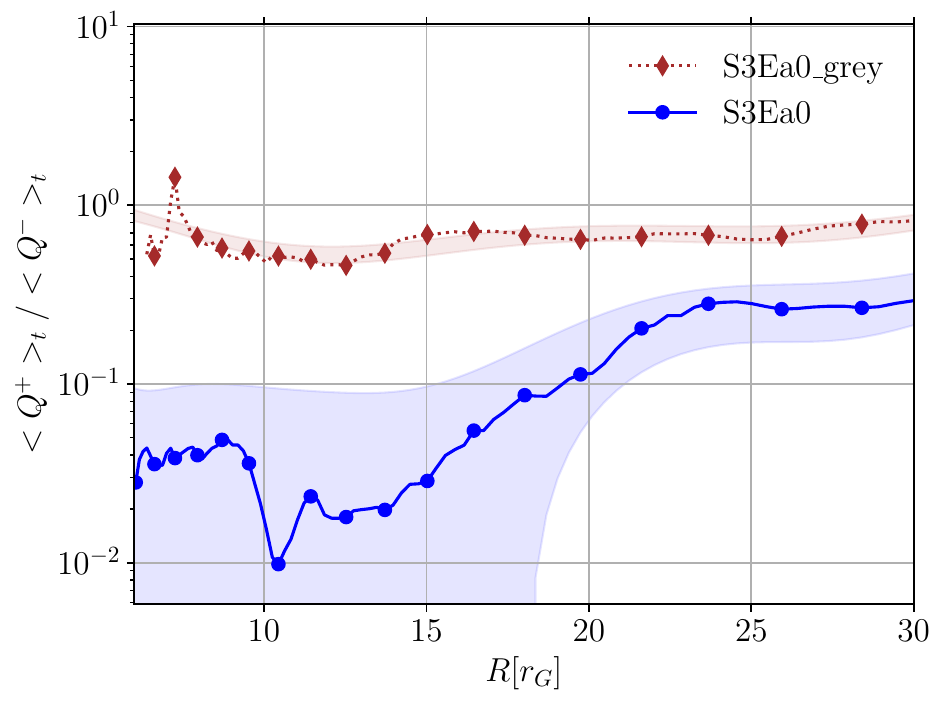}
\includegraphics[width=0.31\textwidth]{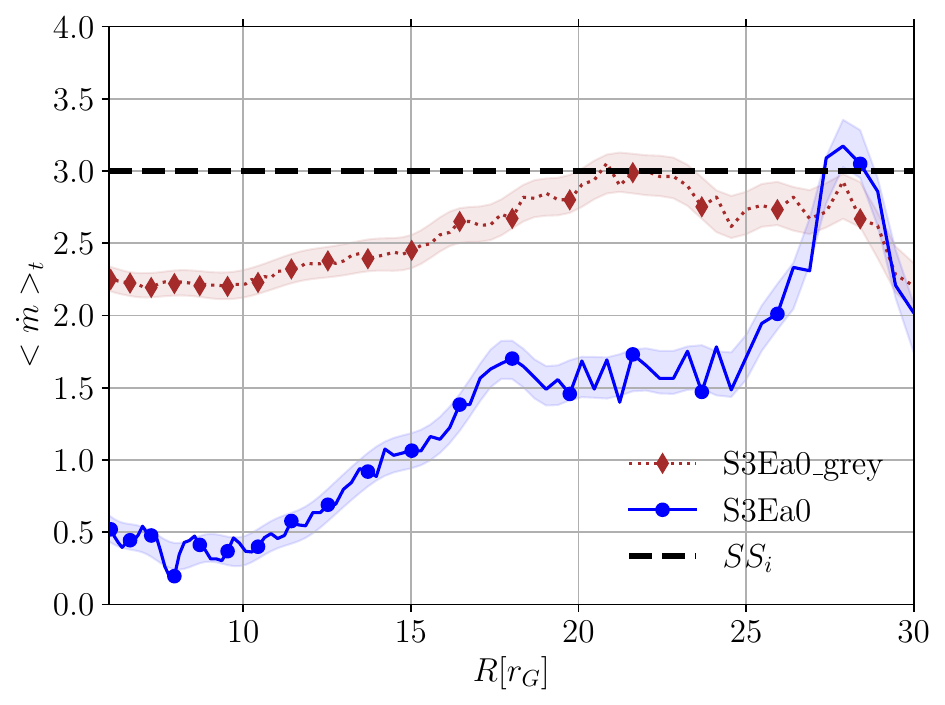}
\caption{Time averaged radial profiles of the disk scale height, $H/R$ ({\em left}); the ratio of heating rate, $Q^+$, to cooling rate, $Q^-$ ({\em middle}); and the mass accretion rate, $\dot{m}$ ({\em right}), for simulations S3Ea0\_grey and S3Ea0. The time averaging is performed over the window $10,000 - 15,000 t_G$. The shaded regions show the $1\sigma$ standard deviations, and the dashed, black $SS_i$ curves show the initial scale height profile and target $\dot{m}$ used for both disks.}
\label{fig:grey-multi}
\end{figure}

\begin{figure}
\centering
\includegraphics[width=0.45\textwidth]{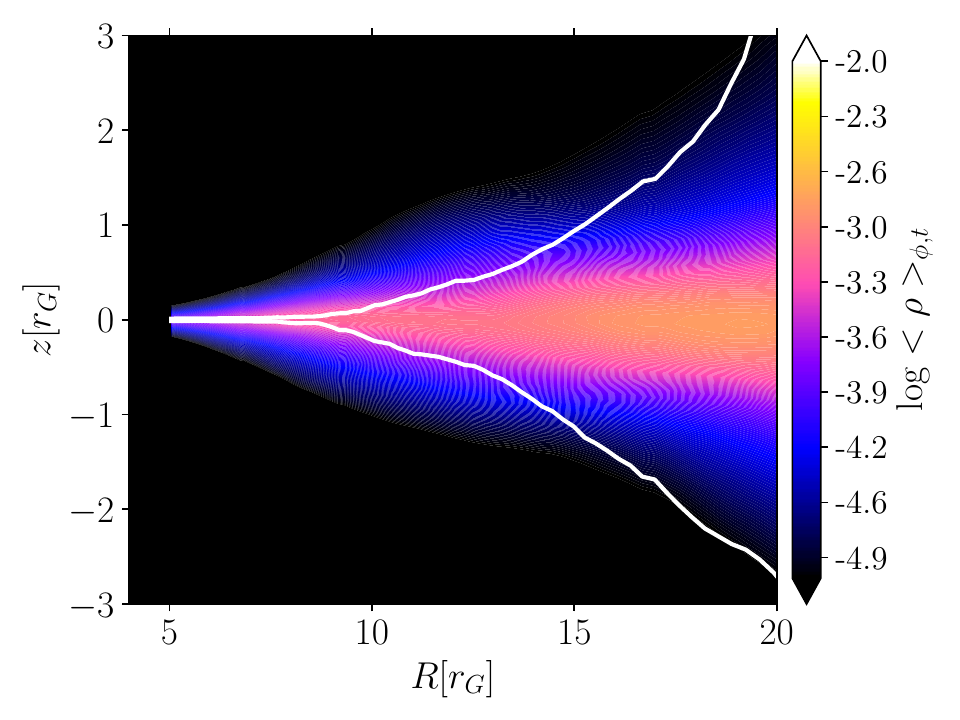}
\includegraphics[width=0.45\textwidth]{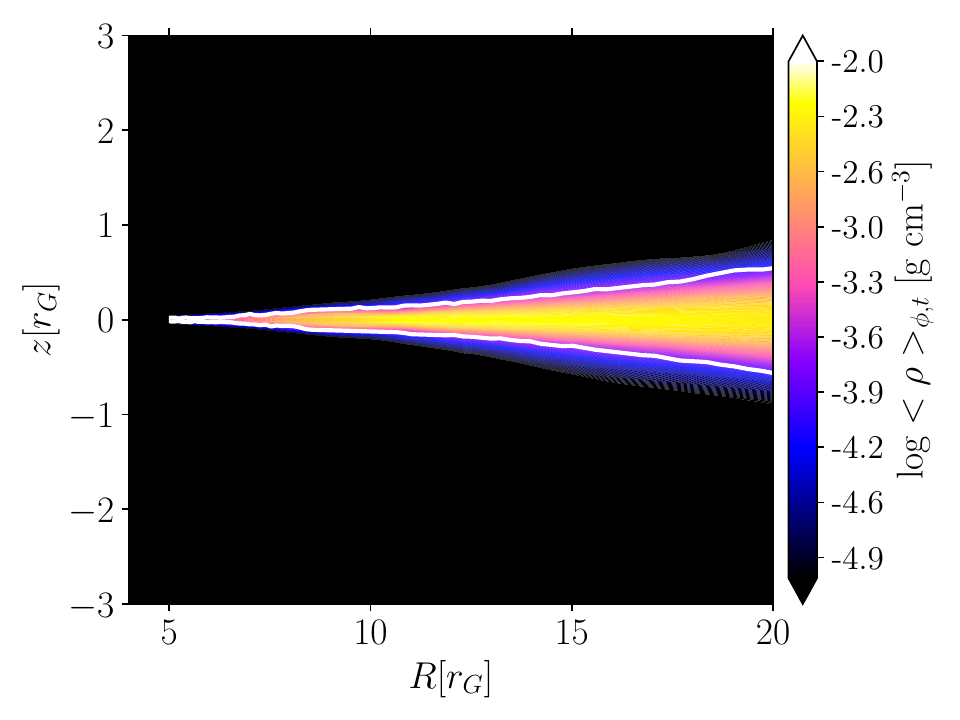}
\caption{Pseudo-color plots of $\langle \rho \rangle_{t,\phi}$ with a white contour showing the effective photosphere for the S3Ea0\_grey ({\em left}) and S3Ea0 ({\em right}) simulations, averaged over azimuth and time for all dumps after $t=6,000 t_G$. Note that $1r_G$ on the vertical axis is stretched compared to $1r_G$ on the horizontal axis. Thus, the image is distorted, which is necessary in order to be able to adequately see the vertical disk structure.}
\label{fig:2D_profiles}
\end{figure}

As we showed in \citet{Mishra22}, the S3Ea0\_grey simulation survives the thermal instability and achieves a quasi-steady state reasonably close to its target configuration (based on the Novikov-Thorne model). S3Ea0 seemingly could not achieve the same. We can see some of the differences in Figure \ref{fig:grey-multi}, where we show time-averaged radial profiles of the two disks. In the first panel, we show the disk scale height, where $H(R)$ is calculated using a density-squared weighting, as
\begin{equation}
\langle H(R)\rangle_\rho = \sqrt{\frac{\int \rho^2 z^2 \mathrm{d}V}{\int \rho^2 \mathrm{d}V}}~,
\label{eqn:height}
\end{equation}
where the integrals are carried out over each radial shell and $\mathrm{d}V$ is the proper volume of a computational element. While the scale height of S3Ea0\_grey actually increased by about a factor of 2, the scale height in S3Ea0 decreased by a similar factor (out to at least $20r_G$). This collapse in the scale height was evidently precipitated by the fact that the heating within the disk was not able to keep up with the cooling, as shown in the middle panel. Notice that for the S3Ea0\_grey simulation, the ratio of heating, $Q^+$, to cooling, $Q^-$, is nearly unity at all radii. In other words, heating roughly balances cooling and the disk is able to maintain thermal equilibrium. This is not the case for S3Ea0, where heating and cooling are far out of balance. To calculate the net heating rate per unit surface area, we take
\begin{equation}
Q^+(R) = \frac{3}{2}\int\langle V^\phi W_{\hat{r}\hat{\phi}}\rangle_\phi \mathrm{d}\theta ~,
\label{heatingrate}
\end{equation}
where the integration is carried out within the limits of the effective photosphere and the integrand is azimuthally averaged, with $V^\phi \approx \Omega$  the azimuthal component of the fluid three velocity and $W_{\hat{r}\hat{\phi}}$  the covariant $r$-$\phi$ component of the MHD stress tensor in the co-moving frame. The radiative cooling is computed by tracking the radiative flux through the effective photosphere\footnote{By computing the radiative cooling at the effective photosphere, we are excluding contributions due to inverse-Compton, which would happen outside this boundary. Thus, the reported cooling rate is a lower bound to the total cooling.} at each radius:
\begin{equation}
Q^-(R) = \langle F^\theta_\mathrm{photo+}(R)\rangle_\phi - \langle F^\theta_\mathrm{photo-}(R)\rangle_\phi~,
\label{coolingrate}
\end{equation}
where $F^\theta_{\mathrm{photo}\pm}(R) = -4/3 E_R u_R^\theta (u_R)_t$ is the flux escaping through the top or bottom effective photosphere, $E_R$ represents the radiation energy density in the radiation rest frame, and $u_R^\mu$ is the radiation rest-frame four-velocity. Both of these calculations require that we track the photosphere of the disk. We do this by integrating the quantity $\kappa_\mathrm{e}\rho$ starting from the pole of the grid and proceeding toward the midplane following a constant coordinate radius, $r$, until $\tau = 1$, where 
\begin{equation}
\tau_< (\theta)= -\int^\theta_{\pi/2}u^t\kappa_\mathrm{e}\rho\sqrt{g_{\theta\theta}}d\theta, \hspace{0.2in} \tau_> (\theta)=\int^\theta_0 u^t\kappa_\mathrm{e}\rho\sqrt{g_{\theta\theta}}d\theta
\label{eq:tau}
\end{equation}
and $\kappa_\mathrm{e} = \sqrt{\kappa^\mathrm{a}_\mathrm{R}(\kappa^\mathrm{a}_\mathrm{R} + \kappa_\mathrm{s}) }$ is the effective opacity. The resulting differences between the two simulations are also illustrated by the 2-D profiles in Figure \ref{fig:2D_profiles}.

\begin{figure}
\centering
\includegraphics[width=0.6\textwidth]{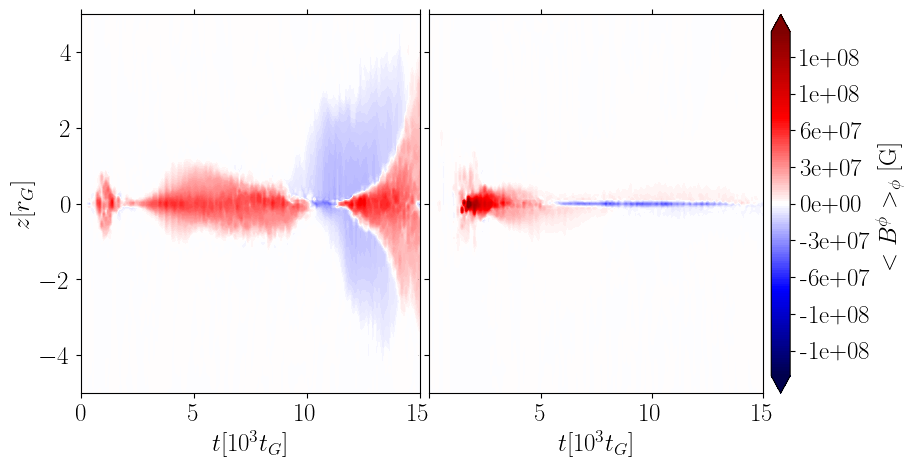}
\caption{Time evolution of the toroidal component of the magnetic field, $B^\phi$, at $R=10\,r_G$ for the S3Ea0\_grey ({\em left}) and S3Ea0 ({\em right}) simulations. The enhanced toroidal field in the grey case provides sufficient vertical support to stabilize the disk, unlike in S3Ea0.}
\label{fig:Bphi}
\end{figure}

As mentioned previously, $\alpha$-viscosity versions of these disks are known to be thermally unstable. What saves the S3Ea0\_grey disk from collapsing is the strong midplane magnetic pressure provided by the toroidal magnetic field component that is continually replenished by its radially extended quadrupole field. The toroidal component is particularly strong near the midplane of the disk, as shown in the left panel of Figure \ref{fig:Bphi}. For some reason, the same build up does not occur in the S3Ea0 simulation (right panel). The relatively weak and collapsed field in the latter simulation is insufficient to support the disk. 

It is a puzzle trying to figure out what happened with the S3Ea0 simulation. On the one hand, it is true that the multi-frequency and grey simulations are likely seeing different cooling rates due to their use of different opacity models. For example, while the grey calculation uses a single power-law for the free-free opacity, the multi-frequency ones use frequency-dependent Gaunt factors that do not exactly integrate up to the grey model. The two methods also use slightly different scattering opacities, $0.325\,\mathrm{cm^2\,g^{-1}}$ for the multi-frequency and $0.34\,\mathrm{cm^2\,g^{-1}}$ for the grey. Finally, the two methods take significantly different approaches to Compton scattering, the most important being that, in the multi-frequency case, the photons can up-scatter to higher frequencies where the effective opacities are lower and can, therefore, more readily escape from the disk, perhaps cooling it more efficiently. We plan to explore these issues further in Roth et al. (in preparation). However, it is certainly not the case that all of our multi-frequency simulations end in collapse. In fact, only S7Ea5 and S3Ea0 failed to recover from the initial relaxation that all of our simulations experienced. Furthermore, as we show in the next section, the mass accretion and luminosity histories of S3Ea0 were initially very similar to the other S3 simulations. So why was it the only one to collapse? The most likely explanation we can come up with is that our simulations are borderline in terms of resolution. If the simulations are right on the edge between stability and instability, then two nearly identical cases, like S3Ea0\_grey and S3Ea0, can diverge in terms of their evolution. Some evidence for this dependence of stability on resolution was presented in \citet{Mishra22}.

\section{Effects of Mass Accretion Rate}
\label{sec:mdot}

\begin{figure}
\centering
\includegraphics[width=0.32\textwidth]{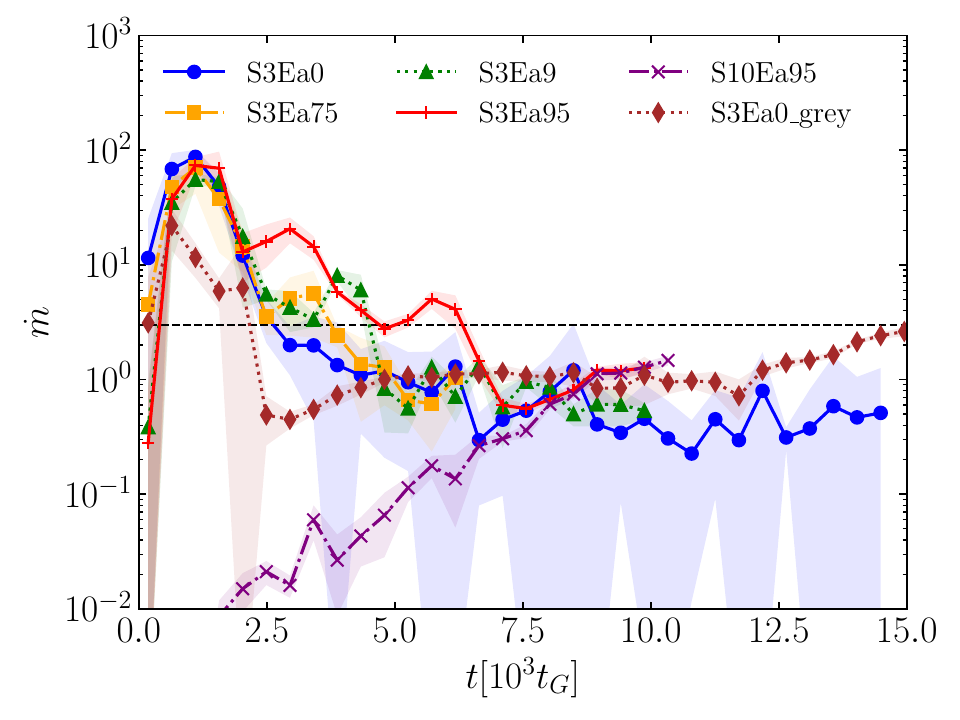}
\includegraphics[width=0.32\textwidth]{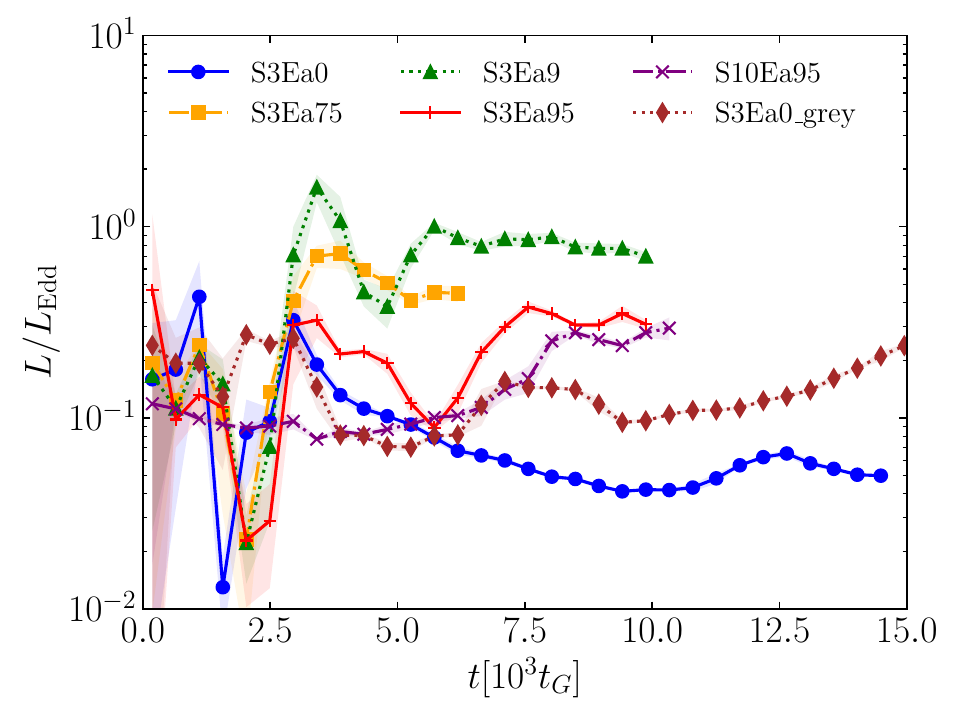}
\includegraphics[width=0.32\textwidth]{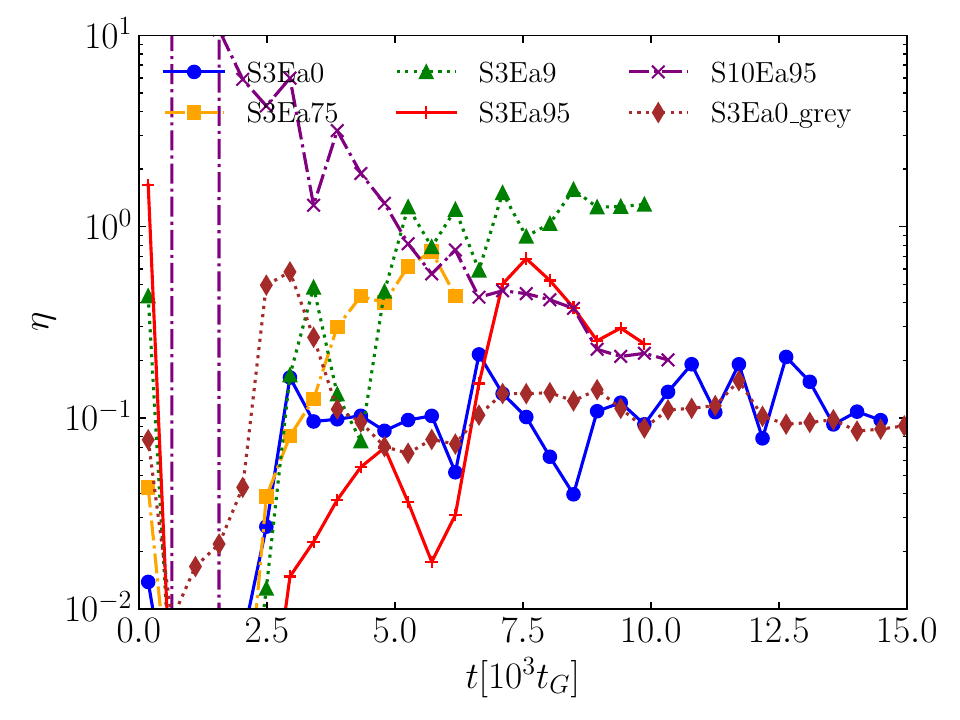}
\caption{{\em Left panel}: The mass accretion rate through the inner boundary of the simulation domain in units of $L_\mathrm{Edd}/c^2$, smoothed using moving averages over five consecutive dumps ($\approx 460 t_G$ in time). The shaded regions show the $1\sigma$ standard deviations, and the black, dashed line shows the target mass accretion rate of $\dot{m}=3$ that applies for all the S3E simulations. {\em Middle panel}: The total luminosity escaping through a cone of $15^\circ$ opening angle at $r=150 r_G$ in units of $L_\mathrm{Edd}$, using the same smoothing as in the left panel. {\em Right panel}: The instantaneous radiative efficiency, $\eta = L/\dot{M}c^2$.}
\label{fig:efficiency}
\end{figure}

To extract $\dot{m}$ from the simulations, we integrate
\begin{equation}
\dot{M}(r,t) = -\int \int \rho u^r \sqrt{-g} {\rm d}\theta {\rm d} \phi
\end{equation}
over a given radial shell and then normalize by $L_\mathrm{Edd}/c^2$. For all of the S3E simulations in Table \ref{tab:params}, we expect the mass accretion rate to be $\dot{m} = 3$ based on the initial Novikov-Thorne profile we used (with $\alpha = 0.02$). As we see in the left panel of Figure \ref{fig:efficiency}, though, all of them start off above this target rate before dropping to quasi-steady values that are factors of $3-6$ below it. As mentioned in Section \ref{sec:methods}, the S7Ea5 simulation had a target mass accretion rate of $\dot{m} = 7$, but it collapsed to a different (much thinner and colder) solution branch, such that its $\dot{m}$ was about a factor of 70 below its target value, and we, therefore, do not include it in Figure \ref{fig:efficiency}. Simulation S10Ea95 is an interesting case, as it is the only one expected to reach super-Eddington luminosities. However, this means that the standard Novikov-Thorne disk solution is not really an appropriate starting point; instead, something like a slim disk \citep{Abramowicz88,Beloborodov98,Sadowski09} may be more appropriate. Thus, it maybe should not be surprising that this simulation started out far from its target mass accretion rate of $\dot{m}=10$ and was one of the few to start well below its target. As this simulation progressed, though, and the disk had time to adjust, $\dot{m}$ increased steadily.

Theoretically, the value of $\dot{m}$ should affect both the luminosity and spectrum of the disk, raising the normalization of each whenever $\dot{m}$ increases. Unfortunately, none of our simulations ended up having sufficiently different values of $\dot{m}$ for us to test this. For instance, simulations S3Ea95 and S10Ea95, which have the same spin and would, therefore, be ideal for comparison, were expected to have mass accretion rates of $\dot{m} = 3$ and 10, but as we mentioned above, S3Ea95 ended up about a factor of 3 below this target, while S10Ea95 started well below its target accretion rate before climbing. As it turned out, the two simulations ended up on nearly overlapping $\dot{m}$ trajectories from $t=7,500 t_G$ onward (left panel of Figure \ref{fig:efficiency}). Given this similarity in $\dot{m}$, it should not be surprising then that the light curves (middle panel) and spectra (Section \ref{sec:spectra}) of these two simulations are nearly identical over the same period. It maybe would have been interesting to see where these two simulations ended up if we had continued to run them. However, because both had issues with their inner boundaries being outside the ISCO, we decided not to pursue them any further than $\approx 10,000 t_G$. We will have to reexamine the effects of mass accretion rate in future work. 


\section{Spectra \& Light Curves}
\label{sec:spectra}

\begin{figure}
\centering
\includegraphics[width=0.45\textwidth]{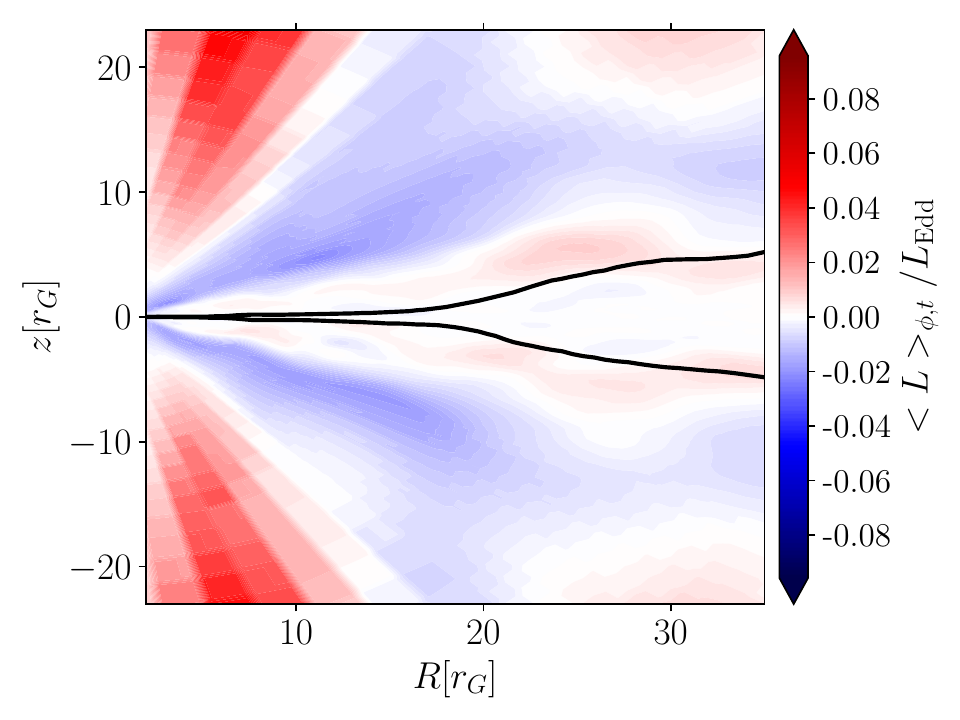}
\caption{Pseudocolor plot showing the local luminosity, integrated over all frequency bins, averaged over azimuth and time for the interval $t=2,500-5,000 t_G$ for simulation S3Ea9. The color scale is chosen such that any radiation moving radially toward the black hole shows up as blue, while outgoing radiation appears red. The solid, black line provides the location of the effective photosphere of the disk. Note that at later simulation times, the radiation in most energy bins is outflowing.}
\label{fig:radiative_flux}
\end{figure}

Perhaps the most exciting aspect of this work is being able to calculate frequency-(or energy-)dependent spectra and light curves directly from the simulations, without having to do a separate radiative transport post-processing step. We found, though, that, in practice, there are some challenges with this. First, the radiation fluxes in the simulations are very dynamic. The radial flux component, $R^r_{t(\nu)}$, can sometimes even be negative, meaning that the radiation is moving towards, rather than away from, the black hole, particularly at early times as shown in Figure \ref{fig:radiative_flux} and more commonly in the lowest energy bins. Furthermore, since the $\mathbf{M}_1$ closure scheme only tracks the net flux of radiation (in a given energy bin) through a given computational zone, it cannot simultaneously track or account for outgoing and ingoing radiation. There are other issues, too, such as the fact that the outer regions of our disks have not had time to reach their equilibrium states. Therefore, it is challenging in any simulations, not just ours, to decide where and how to extract light curves and spectra. Some authors \citep[e.g.,][]{Jiang14,Mishra22} have chosen to track all of the radiation escaping from a radial shell confined relatively close to the black hole ($r \lesssim 20 r_G$), on the logic that this region has most likely reached a near equilibrium state. Others \citep[e.g.,][]{Sadowski16b,Mills23} have chosen to consider radiation escaping through a relatively narrow cone around the symmetry axis, usually far from the black hole, on the logic that this radiation is most representative of what would escape to a distant observer. In this work, we take the latter approach, reporting radiation escaping from within a cone of $15^\circ$ opening angle about the $z$-axis/black hole spin axis at a distance of $150 r_G$ from the black hole (i.e., near the outer boundary of our domain).

\subsection{Spectra}

\begin{figure}
\centering
\includegraphics[width=0.45\textwidth]{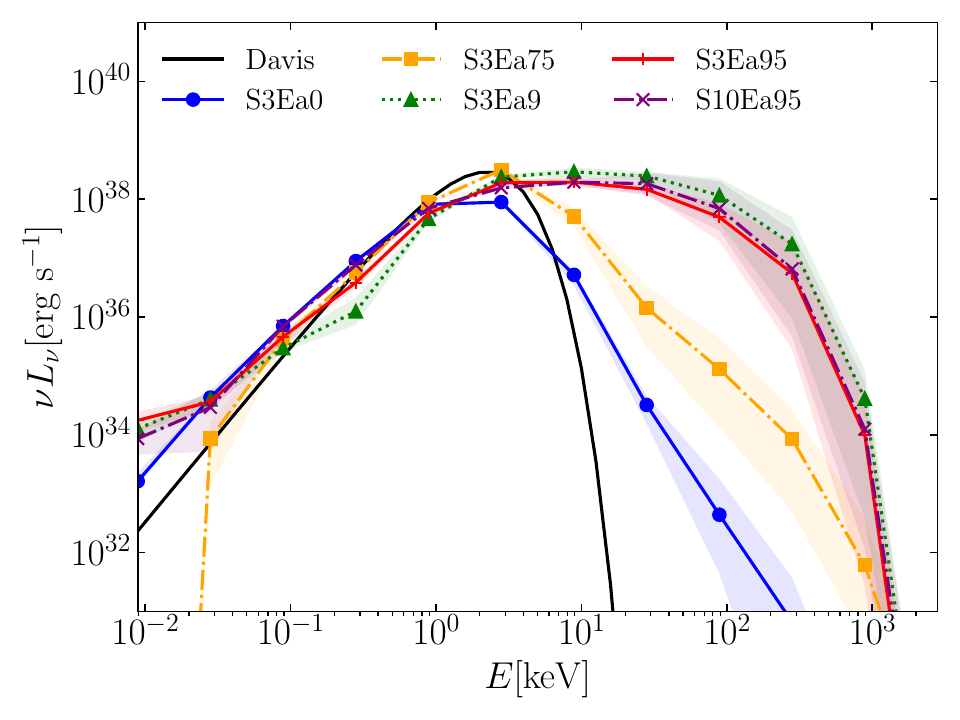}
\caption{Spectra for each simulation accounting for the radiation escaping through a cone of $15^\circ$ opening angle about the $z$-axis at $r=150 r_G$, time averaged over the final $1,000 t_G$ of each simulation, plus a model spectrum from \citet{Shimura95} for $T=3\times10^6$ K and a spectral hardening factor of $f=2.2$. To give some idea of the variability of our spectra, the shaded regions show the minimum and maximum values of the spectra in each bin over the averaging interval. 
}
\label{fig:spectra}
\end{figure}

In Figure \ref{fig:spectra}, we show the spectra extracted from each simulation. To get this, we integrated the radiative flux in each energy bin through a surface at $r=150 r_G$, bounded within cones of $15^\circ$ opening angle about the $z$-axis, as follows
\begin{equation}
L_\nu = -4\left[\int_0^{\pi/2}\int_0^{\pi/{12}} \sqrt{-g} R^r_{t(\nu)} \Delta\nu d\theta d\phi + \int_0^{\pi/2}\int_{11\pi/{12}}^\pi \sqrt{-g} R^r_{t(\nu)} \Delta\nu d\theta d\phi\right] ~,
\label{eq:luminosity}
\end{equation}
where $R^r_{t(\nu)}$ is the radial radiation flux within a frequency bin of width $\Delta \nu$ centered at frequency $\nu$. The factor of 4 in front is to account for the fact that we only simulate a quarter of the full azimuthal domain, and the minus sign is to ensure that outgoing radiation (i.e. $u^r_R > 0$) produces a positive luminosity. The spectra were time averaged over the final eleven dumps ($\approx 1,000 t_G$) of each respective simulation. Figure \ref{fig:spectra} also includes a color-corrected (or diluted) blackbody spectrum \citep{Shimura95} to model the disk, where we have taken the blackbody temperature to be $T=3\times10^6$ K and used a spectral hardening factor of $f=2.2$. This effective temperature is about a factor of 3-6 lower than the actual gas temperatures that we find in our disks. Also, the hardening factor of 2.2 is somewhat higher than what would typically be used for fitting a BHXRB in the thermally dominant, or soft, state \citep[usually $1.4 < f < 2$;][]{Davis19}. This reflects the fact that we find our spectra to all be significantly harder than expected for the soft state, particularly in our higher spin cases. Instead, our spectra are more consistent with the soft intermediate or steep power-law state. This is an issue we explore more in our companion paper (Roth et al. in preparation). 

As mentioned previously, with the $\mathbf{M}_1$ closure method, we find that the radiation in some particular energy bin, at some particular location, and at some particular time, can be inflowing, rather than outflowing. This causes the luminosity in that region to be negative, based on the formal definition in eq. (\ref{eq:luminosity}). For the low-energy ($<0.1$ keV) bins, this is what causes some of the spectra, notably the S3Ea75 one, to drop to the bottom of the plot in Figure \ref{fig:spectra}. For the high-energy ($\gtrsim 10$ keV) bins, we find high, order-of-magnitude or greater, variability, illustrated by the shaded regions in the figure. In between these limits, the portion of the spectrum that can be attributed to the disk, i.e., between 0.1 and 3 keV, is much more stable. 

\subsection{Light curves}

\begin{figure}
\centering
\includegraphics[width=0.45\textwidth]{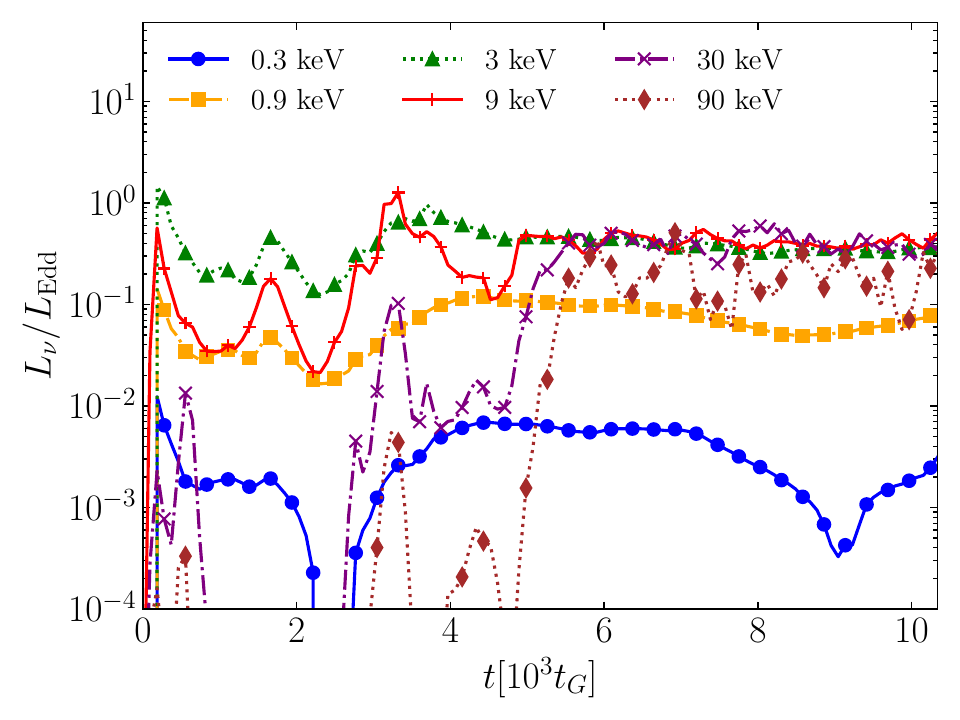}
\includegraphics[width=0.45\textwidth]{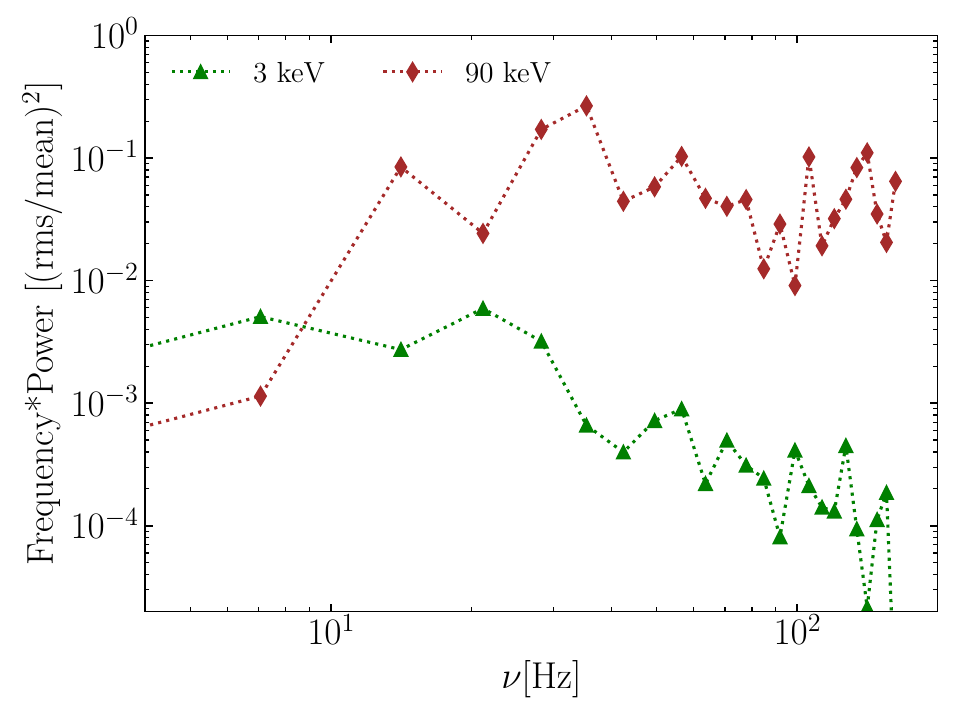}
\caption{{\em Left panel}: Light curves for multiple energy bins for the S3Ea9 simulation, accounting for the radiation escaping through a cone of $15^\circ$ opening angle about the $z$-axis at $r=150 \,r_G$. {\em Right panel}: Power spectra for the 3 and 90 keV light curves over the period from 6,000 to 10,000 $t_G$.}
\label{fig:light_curves}
\end{figure}

Along with spectra, we can also use the simulations to create independent light curves for each energy bin. This is particularly exciting because it gives us a way to track correlated variability between different parts of the disk or the disk and the corona. For instance, the X-ray light curves of BHXRBs have been shown to exhibit frequency-dependent time lags and strong coherence between different energy bands \citep{Miyamoto89,Nowak99,McHardy04}. In previous work, \citet{Bollimpalli20} used $\dot{M}$ measured at different radii in simulations as a proxy for light curves in different energy bins. Even with that crude approach, they found that all of the simulations they considered showed radial coherence and positive time lags (when comparing fluctuations in $\dot{M}$ at smaller radii with those at larger radii) below the viscous frequency. This strongly agrees with what is expected from the inward propagation of fluctuations in the mass accretion rate \citep{Lyubarskii97}. However, \citet{Bollimpalli20} also found that the time lags were frequency-{\em independent}, whereas in X-ray observations they are found to be frequency-{\em dependent}. One way to resolve this discrepancy is to require that the emissivity profile of the disk vary in a certain way with radius. But with our multi-frequency radiation MHD simulations, we do not need to assume an emissivity profile; we can, instead, work directly with light curves in different energy bins such as the ones shown in the left panel of Figure \ref{fig:light_curves}. This is something we plan to do in future work.

With light curves in hand, we can also look at their behavior in frequency space to explore the nature of any variability. The right panel of Figure \ref{fig:light_curves} gives the Fourier power of the 3 and 90 keV light curves for simulation S3Ea9 from $t=6,000$ to $10,000t_G$. Here we are only able to explore a small range of frequencies, limited on the one hand by the time interval between our data dumps, $t_\mathrm{dump} = 92 t_G$, and on the other hand by the length of the time window we are considering. Nevertheless, there are a couple interesting things to note. First, consistent with observations of the soft state \citep[e.g.,][]{Gierlinski05}, we find far more variability in the high-energy (90 keV) component, which is apparent just from looking at the light curves in the left panel. Second, the rms variability near 3 keV is consistent with the variability observed around state transitions in BHXRBs \citep{Done07}. This is another indication that our simulations may match the soft-intermediate state better than the pure soft state. Finally, neither of the power spectra in Figure \ref{fig:light_curves} show significant evidence of excess power at any discrete frequencies, i.e., there is no evidence for quasi-periodic oscillations (QPOs) in this simulation.


\section{Effects of Spin}
\label{sec:spin}

\subsection{Effects of Spin on Spectra}

As mentioned in Section \ref{sec:methods}, we have considered a range of spins from $a_* = 0$ to 0.95. In general, we expect the higher spin cases to exhibit harder spectra for the same $\dot{m}$, simply because the inner radius of the disk, $r_\mathrm{in}$, will move closer to the black hole, leading to a higher maximum temperature. This has been demonstrated in previous work where the radiation was added to the simulations as a post-processing step \citep{Kinch21} and is also what we find. Note that the spectra in Figure \ref{fig:spectra} get progressively harder (the peak moves up and to the right) as the spin increases. 

Not only does the thermal, blackbody part get harder, but we also find more prominent coronal (or hard X-ray) emission in the high-spin simulations. The emission between 10 and 1000 keV increases dramatically as the spin increases. In fact, the radiation in these higher energy bands is probably unrealistically high, likely owing to the approximate treatment of  Compton scattering in our $\mathbf{M}_1$ scheme and perhaps artificially high temperatures near the black hole. This is an issue we plan to explore more fully in our companion paper (Roth et al. in preparation), where we compare the $\mathbf{M}_1$ results to a full Monte-Carlo transport code. 

\begin{figure}
\centering
\includegraphics[width=\textwidth]{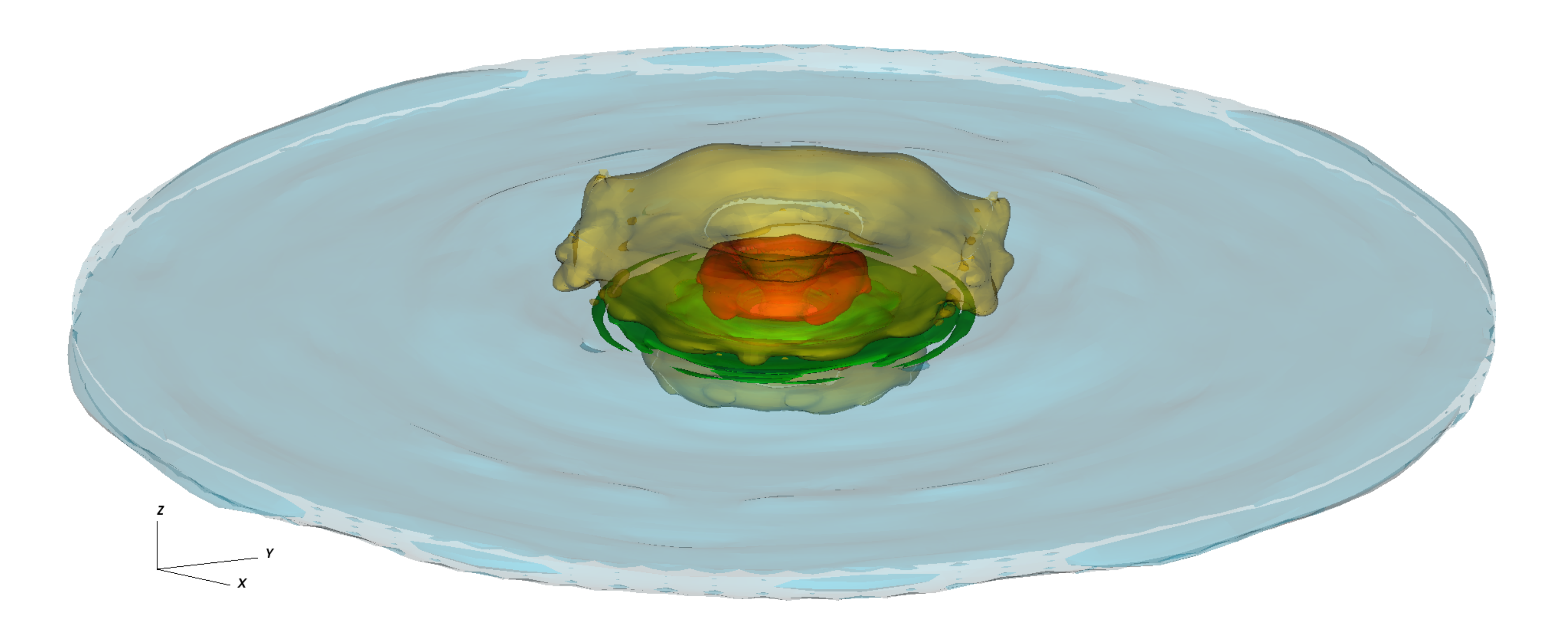}
\caption{Isosurface plots from simulation S3Ea9 at $t=7,900 t_G$ showing roughly where each spectral component dominates the photon energy density. 3 keV emission (blue) is mostly found along the photosphere of the disk; 9 keV emission (green) is concentrated near the inner edge of the disk; 30 keV emission (yellow) is found in a corona sandwiching the inner part of the disk; and 90 keV emission (orange) occupies the hottest regions of the corona surrounding the black hole. For reference, the blue disk in this plot has a radius of $\approx 60 r_G$, i.e., this is not the whole disk nor simulation domain.}
\label{fig:binned_image}
\end{figure}

However, one thing we do not see is evidence for strong X-ray emission from jets. Instead, the geometry of our corona is something in between the doughnut/torus picture and the sandwich corona, as can be seen from Figure \ref{fig:binned_image}. In that figure, we combine four isosurface (isocontour) plots, each for a different energy bin. The contours are based on the radiation energy density within each bin, with the contour level set roughly an order of magnitude below that bin's peak value. In effect, we are plotting where there is a high concentration (high density) of photons of a particular energy. This could be because those photons are thermally generated in that region or because lower energy photons are being Compton up-scattered there. We cannot distinguish this with the information we retain from the $\mathbf{M}_1$ scheme.

The lack of strong jet emission comes down to the fact that the densities and optical depths are extremely low within the funnel region occupied by the jet. The lack of strong emission from the jets should maybe not be considered surprising, since the soft state of BHXRBs, which is the state these simulations are targeting, is not associated with powerful jets \citep{Fender99,Russell11,Maccarone20}. If our simulations are more applicable to the soft-intermediate state, then the role and prominence of jets are less certain.

\subsection{Effects of Spin on Light Curves}

For a given mass accretion rate, higher spins should also lead to higher luminosities. Again, this is because as the spin increases, $r_\mathrm{in}$ moves closer to the black hole, meaning more gravitational binding energy can be extracted before the matter plunges into the black hole. This is generally born out in the middle panel of Figure \ref{fig:efficiency}. Simulations S3Ea0, S3Ea75, and S3Ea9, with spins of $a_*=0$, 0.75, and 0.9, respectively, nicely demonstrate progressively higher luminosities, especially for $t > 3,000 t_G$, once the simulations have started to reach somewhat steady states. The exception to this rule is simulation S3Ea95, with a spin of $a_*=0.95$, where the luminosity is intermediate between S3Ea0 and S3Ea75. However, that simulation is one of the ones that had its inner domain boundary outside the ISCO, which means not all of the gravitational potential energy could be extracted before the matter left the grid.

\subsection{Radiative Efficiency}

Once we have the mass accretion rate and the luminosity, we can calculate a radiative efficiency, $\eta = L/\dot{M}c^2$, for each simulation. In fact, we plot the radiative efficiency as a function of time for all simulations in the right panel of Figure \ref{fig:efficiency}. The most obvious takeaway is that all of our radiative efficiencies are 75-540\% higher than expected for Novikov-Thorne disks at the corresponding spin values (see the fourth column of Table \ref{tab:params} for the Novikov-Thorne values). High radiative efficiencies were also reported in \citet{Kinch21}, ranging from 30-100\% too high for a similar range of spins. Similar to that work, our efficiencies are actually only lower bounds, as we only account for one sixth of the full $4\pi$ steradians in computing our luminosities. The \citet{Kinch21} procedure was also different from ours in that their simulations were non-radiative and instead used an ad hoc cooling prescription. They then post-processed their simulations with a Monte-Carlo radiative transport code. Nevertheless, their study was designed to test thin disks over a similar range of spins and mass accretion rates to ours. In their discussion, they attributed the high radiative efficiencies to cooling happening in the corona, which is not accounted for in the Novikov-Thorne model. That may explain why our radiative efficiencies are even higher, as our $\mathbf{M}_1$ closure seems to produce even more high-energy (coronal) emission than Monte-Carlo radiative transport. We plan to further explore this issue in a companion paper (Roth et al. in preparation), where we reexamine these simulations with our own Monte-Carlo radiative transport code. It is interesting that \citet{Sadowski16b} found radiative efficiencies that were roughly 30-50\% smaller than predicted by the Novikov-Thorne model, though they explored higher (super-Eddington) mass accretion rates.

%

\section{Discussion}
\label{sec:discussion}

We have presented results from six, new, general relativistic, multi-frequency radiation MHD simulations of thin accretion disks for a few different target mass accretion rates and spins. For comparison, we also considered a previously published, frequency-integrated (grey) simulation. The most exciting aspect of this work is being able to generate spectra and energy-dependent light curves directly from the simulations, without special post-processing. This multi-frequency capability may also be crucial to properly modeling intermediate states and state transitions in the future.

Thin accretion disks in the luminosity range we explored ($0.2-1 L_\mathrm{Edd}$) are difficult to simulate for a number of reasons. The simple fact that the disks are thin makes them challenging to sufficiently resolve, but also their infall time is longer than for thick disks, so the simulations ideally need to be run longer, too. Furthermore, these disks are susceptible to a thermal instability that can force them to jump to a different solution branch. Unfortunately, this happened to two of our simulations, S3Ea0 and S7Ea5. S3Ea0 was an illustrative case. Nominally, it should have evolved similar to our earlier, grey-opacity simulation, S3Ea0\_grey, but the two took different paths, the grey one remaining stable, while the multi-frequency one collapsed to a different solution branch. This likely tells us that these simulations are marginally resolved, and under-resolved simulations are more susceptible to the thermal instability than better resolved ones. For S7Ea5, the lower branch solution was so dramatically different that we could not adequately track the evolution of the disk. We, therefore, only presented limited discussions of that simulation.

For the rest of our simulations, we presented time-averaged spectra covering an energy range from $5 \times 10^{-3}$ to $5 \times 10^3$ keV. The resulting spectra were roughly thermal, with peaks around 1-4 keV. Higher energy peaks were associated with higher black hole spins. However, most of the spectra exhibited much greater power at energies above 10 keV than expected. This emission is mostly associated with very hot gas in our coronae, which may be a consequence of the approximate treatment of Compton scattering in our $\mathbf{M}_1$ closure method or our numerical procedures for handling the highly magnetized gas near the black hole. We discuss below our plans to explore this further in future work.

Despite the high temperatures of our coronae, we still feel it is worthwhile to remark on their geometries. We found them to be roughly doughnut shaped, filling in the regions interior to the disks, but also sandwiching the disks out to $10-20 r_G$. We did not see evidence for strong jets, nor X-ray emission associated with any jets, even in our highest spin ($a_*=0.95$) cases.

We also presented energy-dependent light curves and corresponding power spectra. Consistent with the harder-than-expected spectra, we found that the high-energy ($>10$ keV) light curves were brighter and more variable than expected from real BHXRBs in the soft or soft-intermediate states. The variability near the thermal peak, though, was roughly consistent with power spectra of the soft-intermediate state.

Our simulations did not produce a broad enough range of $\dot{m}$ values to see any effect of this variable on our spectra or light curves. This is something we will have to try again to explore in future work.

Finally, we tested a few different black hole spins. There we did find differences in both the spectra and light curves. As expected, higher spins led to harder spectra (both a higher thermal peak, but also a much harder spectral tail) and generally brighter luminosities. In fact, the luminosities are higher than expected for the observed mass accretion rates, leading to radiative efficiences that are factors of at least a few higher than the corresponding Novikov-Thorne values. This has been seen in previous, comparable work \citep[e.g.,][]{Kinch21}, and may be attributed to the strong coronal emission found in each of our simulations.

\subsection{Future Work}

Now that we can create light curves for each energy bin independently (instead of just a single, summed light curve), we have an opportunity to look for correlated variability between different bins, much the way observers do. We saw signs of correlated variability in numerical simulations of black hole accretion disks using the mass accretion rate at different radii as a proxy for light curves in different energy ranges \citep{Bollimpalli20}, but now we can do direct cross-correlations of the separate light curves. 

Additionally, we plan to further analyze our light curves in Fourier space to look more carefully for QPOs. Again, the binning of our radiation by energy will allow us to look for QPOs in the different spectral components and then map out the specific locations within the accretion flow where the oscillations originate, possibly pointing us to their physical origins. QPOs are not prominent in the soft state, but often reach their peak power in the intermediate states, which is where we will focus our future work.

Finally, the $\mathbf{M}_1$ method used in this paper treats Compton scattering by solving the Kompaneets equation, which is valid in the small energy limit and approximates scattering as a diffusion operator up and down frequency bins. We generally expect this approach to give broader spectra than a full treatment of Compton scattering, especially in the high-spin cases, as the temperatures in those simulations, especially near the black hole, reach values that are at the edge of validity for Kompaneets. In our companion work (Roth et al. in preparation), we compare the results of our $\mathbf{M}_1$ closure scheme with a full, Monte-Carlo radiation transport scheme that has a proper, relativistic (hot) treatment of Compton scattering and does not approximate it as a diffusion process. In that work, we explore the relative contributions that these two issues, high temperatures and the diffusion approximation, have on the spectra. 

\begin{acknowledgments}
We thank the referee for their useful feedback on this manuscript. P.C.F. gratefully acknowledges support from the National Science Foundation under grants AST-1907850 and PHY-1748958. This work was performed in part at the Aspen Center for Physics, which is supported by National Science Foundation grant PHY-1607611. The work by P.A. and N.R. was performed under the auspices of the U.S. Department of Energy by Lawrence Livermore National Laboratory under Contract DE-AC52-07NA27344. This work used the Extreme Science and Engineering Discovery Environment (XSEDE), which is supported by National Science Foundation grant number ACI-1053575.
\end{acknowledgments}

%

\vspace{5mm}
\facilities{XSEDE(Stampede2)}


\software{Cosmos++ \citep{Anninos05,Fragile12,Fragile14,Anninos20}}





\bibliographystyle{aasjournal}



\end{document}